\documentclass[fleqn,12pt,twoside]{article}
\usepackage{espcrc1}

\usepackage{graphicx}
\usepackage[figuresright]{rotating}

\newcommand{\AmS}{{\protect\the\textfont2
  A\kern-.1667em\lower.5ex\hbox{M}\kern-.125emS}}

\title{A precise method for the detection of the two photons from the $\pi^0$ decay in hypernuclear spectroscopy}

\author{T. Petkovi\'c \address[MCSD]{Department of Applied Physics, Faculty of Electrical Engineering and Computing,\\University of Zagreb, Croatia}%
        and
        D. Hrupec\addressmark
       }
       
\begin{document}

\maketitle

\begin{abstract}
An experimental technique for the light $\Lambda$-hypernuclei structure studies by using $(K^-,\pi^0)$ reaction and Neutral Meson Spectrometer (NMS) developed at the BNL has been described. Position dependence
calibration of the BGO conversion planes of the NMS was invented as the solution to the crucial constraint of
the high resolution in single-$\Lambda$ spectroscopy. Position parameters of the BGO crystal rods were fitted out from the out-of-kaon beam measured data as obtained by the original small highly collimated $^{60}$Co source method, based on the coincidences between top and bottom signals of each BGO rod.
\end{abstract}

\section{Introduction}

A scientific connection of hypernuclear to nuclear few-body physics seems to be well established by the modern picture that $\Lambda$ particle plays \textit{a glue like role} in hypernucleus. According to that picture $\Lambda$-particle acts towards the centre producing a dynamical contraction of the core nucleus. Such picture opens \textit{per definitionem} interesting issues of the effects of $\Lambda$-Nucleon ($\Lambda N$)
spin-orbit force as well as Hyperon-Hyperon ($YY$) interaction, both for the single-$\Lambda$
and double-$\Lambda$ hypernuclei. This is a basic direction of research in contemporary hypernuclear
physics dealing with $\Lambda$ and $\Sigma$ hyperons ($S=-1$), and broadening to
a higher strangeness degree of freedom by including of $\Xi$ hyperons ($S=-2$). The crucial steps
have been recently made by the Japanese groups of Y.~Akaishi \cite{Nemu02} and of E.~Hiyama \cite{Hiya02}.
Within \textit{ab initio} variational calculations of the s-shell single-$\Lambda$ hypernuclei,
the important role was attributed to the $\Lambda N$-$\Sigma N$ coupling which strongly affects an internal structure of the core nucleus \cite{Nemu02}. In case of double-$\Lambda$ hypernuclei, beyond the Pauli suppression
effect the $\Lambda\Lambda$-$\Xi N$ coupling effect was investigated to be significant for the proper
evaluation of the $YY$ interaction \cite{Hiya02}. Our understanding of the $YN$ and $YY$ interactions may be searching for within the two theoretical frameworks: meson theory (meson exchange model) which is based on the modern potentials of few-body physics and/or quark model. Which way of answering will be conceptually more or less accepted,
depends exclusively on the present and near future hypernuclear experiments. At present, two state-of-the-art experimental techniques on $\Lambda$-spectroscopy seem to dominate in hypernuclear physics. The first, based on the reaction $(K^-_{stopped},\pi^0)$ and Neutral Meson Spectrometer, was developed by the E907 and E931 experiments at the BNL. In both experiments, hypernuclei were created by producing $\Lambda^0$ via the strangeness exchange reaction by using secondary beam of stopping kaons. Another high resolution technique was developed at the \textit{JLab} by the \textit{HNSS Collaboration}. That is the electroproduction of $\Lambda$-hypernuclei by the $(e,e'K^+)$ reaction. In this paper, our attention is focussed on the first method which is, essentially, based on the NMS and its detection properties and parameters.

\section{Experiment and Results}

A Neutral Meson Spectrometer (NMS) constructed and commissioned at LAMPF, has been used
by the Collaboration Experiment E907 \cite{Ahme01} at the Alternating Gradient Synchrotron (AGS) of the
BNL for $\Lambda$-hypernuclear spectroscopy based on the reaction: 
$K^{-}+{}^{A}Z\longrightarrow{}_\Lambda^{A}(Z-1)+ \pi^0$,
when a recoiling hypernucleus and a $\pi^0$ are produced. The E907 was particularly dedicated for
the reaction:
$K^{-}+{}^{12}C \longrightarrow{}_\Lambda^{12}B + \pi^0$,
in order to study s- and p-level states in the excitation spectrum of the
$_\Lambda^{12}B$ hypernucleus. Reaction vertex reconstruction is well determined by the $\pi^0$ production and its immediate electromagnetic decay ($\tau = 0.83\cdot10^{-16}\,{\rm s}$) in two photons, without loosing energy in the target at the $\Lambda$-production point. A high spectroscopic resolution is thoroughly defined by the kinematically complete measurement of two photons from the $\pi^0$ decay:   
$$\pi^0\to\gamma\gamma,\quad
x={E_1-E_2\over E_1+E_2},\quad{\rm and}\quad
E_{\pi}(x,\eta)=m_{\pi}\sqrt{2\over(1-\cos\eta)(1-x^2)}$$
where $x$ is energy sharing parameter between two photons, $\eta = \eta_{lab}$ is opening angle between
two photons as seen by the \textit{Arm1} and \textit{Arm2} of the NMS, and $E_{\pi}$ is total energy of the $\pi^0$
to be reconstructed from the two photons. A schematic view of NMS including large area veto scintillation counter, BGO conversion planes, MWPCs, and CsI crystal array of each arm can be seen in ref.\ \cite{Ahme01,Hrup01}.
Within the frame of the E907 engineering runs off-line data analysis, we found the BGO-layers
are very critical and sensitive parts in view of the overall energy resolution of the NMS.
Fig. 1. shows a conceptual front view of the single BGO conversion plane. It consists of 14 rods of $5\times1\,{\rm cm}^2$ cross section and 40 cm length each. The individual rod was composed from the two identical bars of
20 cm length, coupled together and viewed from the ends by photomultipliers (\textit{PMTs}). It can be proved by the Taylor's derivative expansion of the $E_{\pi}(x,\eta)$ function that uncertainty in $\pi^0$ energy, arising from the BGO thickness ($\Delta x = 1\,{\rm cm}$) and opening angle ($\eta = 52^{\circ}$) of the NMS real set-up, should be less than 0.02 MeV \cite{Hrup01}. Hence, a top-bottom position dependence along the BGO rod length only remained to be investigated. We developed a precise method for calibration of the two conversion BGO front planes in each arms of NMS, which convert photons from the $\pi^0$ decays into showers. Gain equalization and position dependence of the output pulses (voltage-time area, in units of \textit{nVs}) over an entire length of the each rod were calibrated by the small collimated $^{60}$Co source coincident method. A method was based on acquiring coincidences between top and bottom signals of the individual BGO rod, while a collimated source was located at the particular place of the five different positions ($x/{\rm cm} = 8, 14, 20, 26, 32$, respectively) along $40\,{\rm cm}$ length of the crystal. A basic function used to figure out a position dependence of BGO has the form of
$I(x)=I_0e^{-x/\lambda_{eff}}$. Effective attenuation length $\lambda_{eff}$ was fitted out for each top-bottom BGO rod, which finally gave a column of 112 calibration values which correspond to four conversion planes of the NMS.
A mean value of $\bar\lambda_{eff} = 91.6\,{\rm cm}$ was determined on the set of all data. Variation in the intensities of the pulse comparing a position at the near edge with respect to the middle of the BGO rod was evaluated to be $\simeq 2 \%$, and should be ignored assuming $\bar\lambda_{eff}$ as an overall calorimeter parameter for the entire BGO plane. However, the strongest attenuation length of $\lambda_{eff} = 35\,{\rm cm}$ was evaluated from the measured data \cite{Hrup01}. For that case, a variation in the intensities was calculated to be $14.4\%$, and hence to be used for correction of the NMS overall energy resolution. In Fig. 2. a position dependence of the individual BGO rod is shown.

\section{Conclusions}

By the NMS a reaction $(K^-_{stopped},\pi^0)$ became a new tool for the high resolution $\Lambda$-spectroscopy. Our hypothesis of position dependence of the BGO-layers has been clearly demonstrated by the small collimated $^{60}$Co source coincident method. A significant variations in $\lambda_{eff}$ of individual rods were figured out. The position sensitive parameters of the four BGO conversion planes of the NMS have inevitably to be integrated in the data analysis program of the E907.



\begin{figure}[htb]
\begin{minipage}[t]{78mm}
\includegraphics[height=4cm]{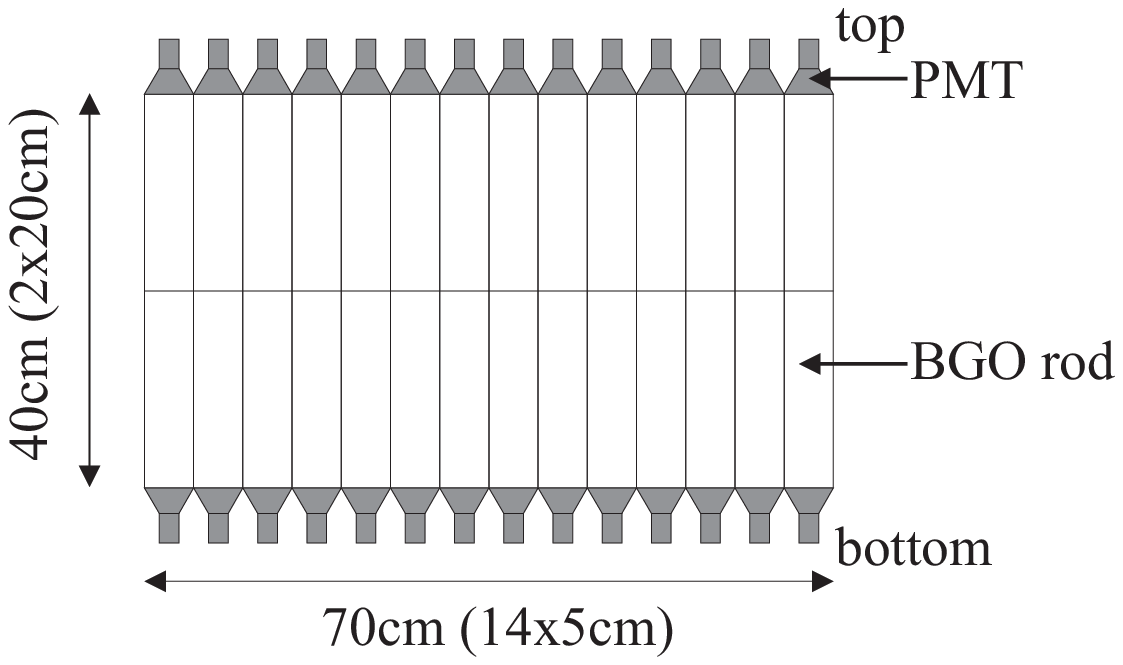}\vskip -10mm
\caption{Front view of the BGO conversion plane of the NMS.}
\label{fig:assembly}
\end{minipage}
\hspace{\fill}
\begin{minipage}[t]{78mm}
\includegraphics[height=4cm]{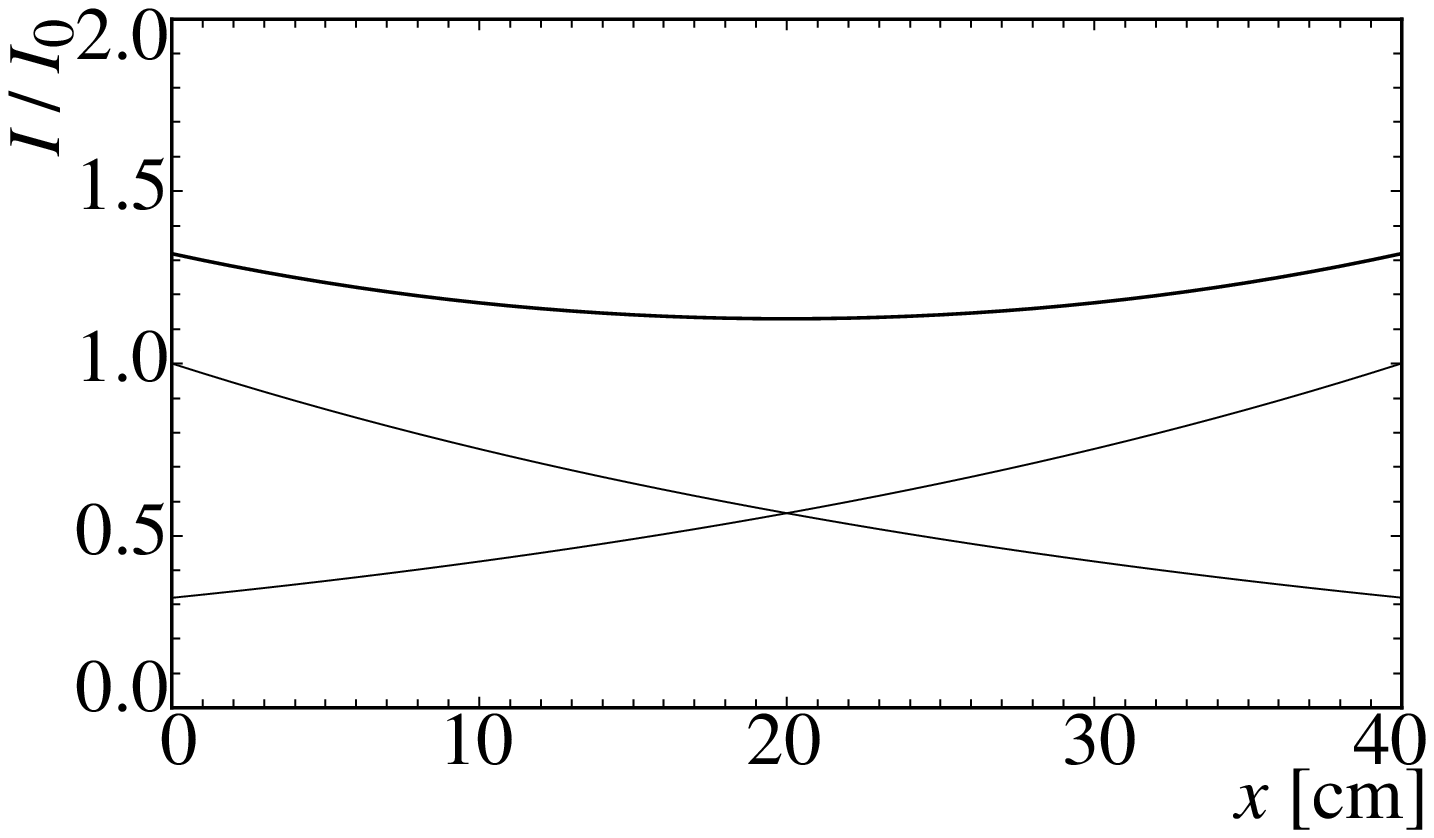}\vskip -10mm
\caption{Position dependence of the BGO crystal rod of the NMS.}
\label{fig:positiondependence}
\end{minipage}
\end{figure}

\end{document}